\documentclass[a4paper,onecolumn, 11pt]{quantumarticle}
\pdfoutput=1
\usepackage[utf8]{inputenc}
\usepackage[english]{babel}
\usepackage[T1]{fontenc}
\usepackage{amsmath}
\usepackage{hyperref}
\usepackage{tikz}
\usepackage{lipsum}

\usepackage{amssymb}

\begin{document}

\title{Norm Inequality for Perturbed Quantum Evolutions and Its Application to 
Grover’s Algorithm}

\author{Kohei Kobayashi}
\email{k\_kobayashi@nii.ac.jp}
\affiliation{
Global Research Center for Quantum Information Science, National Institute of Informatics, 2-1-2, Hitotsubashi, Chiyoda-Ku,Tokyo 101-8430, Japan
}
\orcid{0009-0004-5392-6744}


\maketitle

\begin{abstract}
  We investigate the impact of coherent control errors on quantum state evolution 
by deriving a general norm inequality based on Grönwall’s lemma. 
This inequality provides an explicit upper bound on the deviation between an ideal quantum state and one subject to arbitrary coherent perturbations, including both time-dependent and time-independent cases. 
The framework is broadly applicable, requiring no assumptions 
about the detailed structure of the perturbation and full dynamics of the quantum system. 
We apply this approach to analyze the robustness of Grover’s search algorithm in the presence of coherent errors.
Our results characterizes quantitative scaling relations between the error strength, 
algorithm runtime, and success probability, 
offering practical guidelines for the design of resilient quantum protocols. 
We further compare the effects of time-dependent and time-independent perturbations, 
showing the distinctive ways in which coherent errors accumulate in quantum dynamics.
\end{abstract}

\section{Introduction}

The performance of quantum algorithms is typically analyzed under the assumption of ideal unitary dynamics \cite{Nielsen}. 
In practice, however, quantum systems are inevitably subject to various imperfections in control operations, 
including systematic coherent errors. 
These errors arise from inaccuracies in control fields, miscalibrations of pulse amplitudes or phases, 
and residual couplings to unwanted degrees of freedom. 
Their impact can be particularly severe in algorithms that rely on precise interference patterns, 
such as Grover’s search algorithm \cite{Grover, Farhi, Roland, Biham}.

While incoherent errors have been extensively studied \cite{error1, error2, error3, error4, error5, error6}
  and can often be mitigated 
by error correction \cite{QEC1, QEC2} 
or noise-averaging techniques \cite{DD1, DD2}, 
coherent errors pose a more subtle challenge. 
Because they are systematic and accumulate coherently over time, 
their effects can be difficult to diagnose and correct. 
A quantitative understanding of how coherent errors affect quantum state evolution is 
therefore essential for the reliable implementation of quantum algorithms.

In this paper, we provide a rigorous and general framework for analyzing the 
influence of coherent control errors on isolated quantum dynamics. 
By comparing the evolution of a quantum system under an ideal Hamiltonian with that under a perturbed Hamiltonian, 
we derive a Grönwall-type inequality that bounds the norm of the deviation between the ideal and actual states. 
This inequality is applicable to arbitrary time-dependent and time-independent coherent perturbations, 
without requiring specific assumptions about their structure.

We further apply this framework to assess the robustness of Grover’s algorithm in the presence of coherent errors. 
Our analysis clarifies how the success probability of the algorithm depends quantitatively on the error strength and evolution time. 
In particular, we identify scaling relations that characterize the tolerable levels 
of coherent error for maintaining high-fidelity algorithmic performance.

\section{Grönwall-type bound on state deviation}

We consider two quantum states $|\psi(t)\rangle$ and $|\phi(t)\rangle$
evolving from a common 
initial states $|\psi(0)\rangle=|\phi(0)\rangle=|\psi_0\rangle$ under two different Schrödinger equations:

\begin{align}
\label{ideal}
\frac{d}{dt}|\psi(t)\rangle &= -i H(t) |\psi(t)\rangle, \\
\frac{d}{dt}|\phi(t)\rangle &= -i \left( H(t) + K(t) \right) |\phi(t)\rangle, 
\end{align}
where \( H(t) \) is the ideal (Hermitian) Hamiltonian, and \( K(t) \) represents a 
time-dependent perturbation or control error (we set $\hbar=1$ in the following).
Although $K(t)$ can in principle be non-Hermitian, 
we assume throughout this paper that $K(t)$ is Hermitian, 
consistent with the assumption of an isolated quantum system.

 Let us define the deviation vector 
 $|\delta(t)\rangle:= |\psi(t)\rangle - |\phi(t)\rangle$.
  Taking the time derivative and using Eqs. (\ref{ideal}) and (2), we obtain
\begin{align}
\label{deltaevolution}
\frac{d}{dt}|\delta(t)\rangle 
&= -i H |\psi(t)\rangle + i \left(H + K(t)\right) |\phi(t)\rangle \nonumber \\
&= -i H |\delta(t)\rangle + i K(t) |\phi(t)\rangle. 
\end{align}

We now derive a bound on the Euclidean norm \( \|\delta(t)\|=\sqrt{\langle \delta(t)|\delta(t)\rangle} \).
We consider a time derivative
\begin{align}
\label{deltaevolution2}
\frac{d}{dt} \| |\delta(t)\rangle \|^2 
&= 2\Re \left\{ \langle \delta(t)|\frac{d}{dt}|\delta(t)\rangle \right\} \nonumber \\
&= 2\Im \left\{ \langle \delta(t)| K(t) |\phi(t)\rangle \right\} \nonumber \\
&\leq 2\| |\delta(t)\rangle \| \cdot \|K(t) |\phi(t) \rangle \|  \nonumber \\
&\leq 2\| |\delta(t)\rangle \|\cdot  \|K(t)\| \cdot \| |\phi(t)\rangle \|,
\end{align}
where we used the Schwarz's inequality and $\|A\|$ is the operator norm.
On the other hand, 

\begin{eqnarray}
\label{deltaevolution3}
\frac{d}{dt} \| |\delta(t)\rangle \| ^2=2\| |\delta(t)\rangle \| \frac{d}{dt} \| |\delta(t)\rangle \|.
\end{eqnarray}
Therefore, we obtain
\begin{eqnarray}
 \label{simplebound}
\frac{d}{dt}\|\delta(T)\| \leq  \|K(t)\| \cdot \| |\phi(t)\rangle \|.
\end{eqnarray}

From Eq. (\ref{deltaevolution}), we compute:

\begin{align}
\label{gronwall-pre}
\frac{d}{dt} \|\delta(t)\| &\leq \|K(t)\|  \left( \|\psi(t)\| + \| |\delta(t)\rangle \| \right)  \nonumber \\
&\leq \|K(t)\| \left(1 + \||\delta(t)\rangle \| \right). 
\end{align}

This differential inequality is of the form:
\begin{eqnarray}
\frac{d}{dt} f(t) \leq \kappa(t)\left( 1 + f(t) \right),
\end{eqnarray}
where $f(t):= \||\delta(t)\rangle\|$ and  $\kappa(t):=\|K(t)\|$.
Then, 
\begin{eqnarray}
\frac{d}{dt} \left( e^{-\int^s_0\kappa(t)dt} f(t)\right) \leq \kappa(t)e^{-\int^T_0\kappa(s)ds},
\end{eqnarray}

we obtain the bound:
\begin{eqnarray}
\label{inequality}
\| |\psi(T)\rangle- |\phi(T)\rangle \| 
\leq \int_0^T \|K(s)\| \exp\left( \int_s^T \|K(u)\| du \right) ds. 
\end{eqnarray}

This result reveals how the deviation accumulates over time 
by taking into account both the instantaneous strength of the perturbation and its future amplification potential.
In particular, it shows that early-time errors can be exponentially magnified, 
which gives us the importance of suppressing even transient control imperfections.

This bound is also useful from a practical point of view, 
as it provides an estimate of how much the perturbed state may 
deviate from the ideal state without requiring a full simulation of the system’s time evolution. 
The bound depends only on the norm \( \|K(t)\| \) of the error term, which can often be calculated or estimated from experimental parameters.

Moreover, the bound involves only integrals over simple scalar functions, making it numerically efficient even for large quantum systems.
The presence of the exponential factor also represents how early-time errors can have a disproportionate impact on the final state, offering both a worst-case estimate and valuable insight into the temporal sensitivity of the protocol.

We note that the Grönwall-type inequality derived here 
formally applies to general time-dependent operators \( K(t) \), including non-Hermitian cases. 
However, in this paper we focus exclusively on isolated quantum systems, and \( K(t) \) is assumed to be Hermitian throughout the analysis.
Extending this approach to open quantum systems 
and general non-Hermitian errors is an interesting direction for future research.

\section{Analysis of  success probability}

Grover's search algorithm provides a quadratic speedup over classical search one for identifying a marked item 
in an unstructured database of \( N \) elements \cite{Grover}.
The algorithm begins in the equal superposition state 
\( |+\rangle = (1/\sqrt{N}) \sum_{x=0}^{N-1} |x\rangle \), 
and iteratively applies a sequence of unitary operations that rotate the state toward the marked solution \( |w\rangle \).
The ideal dynamics can be viewed as a rotation in a two-dimensional subspace spanned by \( |w\rangle \) and its orthogonal complement,
 with an effective Rabi frequency $\Omega = 2\theta \approx 2/\sqrt{N}$. 
 After $T \approx\pi/(2\Omega)= \pi \sqrt{N}/4$ time units, 
 the state is nearly aligned with \( |w\rangle \), achieving success probability close to one.

To study the effect of control errors, 
let us consider the resulting degradation in the success probability due to the perturbation.
Assuming that the ideal state at the final time \( T \) satisfies \( |\psi(T)\rangle \approx |w\rangle \), 
we define the success probability of the perturbed protocol as
\begin{equation}
P_{\text{succ}} := |\langle w|\phi(T)\rangle|^2.
\end{equation}
Using the triangle inequality, we obtain a lower bound on the success probability in terms of the state deviation norm:
\begin{align}
\label{eq:succ-bound1}
|\langle w|\phi(T)\rangle| 
&\geq |\langle w|\psi(T)\rangle| - \| |\psi(T)\rangle - |\phi(T)\rangle \| \nonumber \\
&\approx 1 - \| |\psi(T)\rangle - |\phi(T)\rangle \|. 
\end{align}

Substituting the Grönwall-type inequality derived in Eq. (\ref{inequality}), we arrive at
\begin{equation}
\label{eq:succ-bound2}
P_{\text{succ}} 
\ge \left( 1 - \int_0^T \|K(s)\| \exp\left( \int_s^T \|K(u)\| du \right) ds \right)^2. 
\end{equation}

This inequality quantifies how the strength and structure of the perturbation \( K(t) \) 
influence the success probability of algorithm.
Notably, early-time errors may have a stronger impact on the final outcome due to the exponential weighting.
If the operator norm \( \|K(t)\|\) is uniformly bounded 
by a constant \( \|K(t)\|\leq \gamma \), 
then Eq. (\ref{eq:succ-bound2}) reduces to
\begin{equation}
P_{\text{succ}} \ge \left(1 - \left( e^{\kappa T} - 1 \right) \right)^2 = \left(2 - e^{\kappa T} \right)^2. \label{eq:succ-bound-const}
\end{equation}
In the regime \( \kappa T \ll 1 \), 
we have \( e^{\kappa T} \approx 1 + \kappa T \), and the bound simplifies to
\begin{equation}
P_{\text{succ}} \gtrsim 1 - 2\kappa T + \mathcal{O}((\kappa T)^2).
\end{equation}

This provides a clear quantitative condition for error tolerance; that is,
 the integral of the error strength over the protocol duration must be sufficiently small to ensure high-fidelity operation.
 As such, the Grönwall-type bound not only yields a rigorous estimate of algorithm performance 
 under coherent errors but also offers a useful criterion for designing error-resilient control strategies.

To guarantee that the success probability remains above a fixed threshold \( 1-\epsilon \),
the error strength \(\gamma \) must satisfy a constraint determined by the total runtime.
In Grover’s algorithm, the optimal runtime scales as \( T=\pi \sqrt{N}/4 \),
where \( N \) is the problem size. 
Substituting this into the deviation bound yields
\begin{equation}
\gamma \leq \frac{4}{\pi \sqrt{N}} \ln\left(2 - \sqrt{1 - \epsilon}\right).
\end{equation}
In the small \(\epsilon\) limit, this reduces to 
\( \gamma \lesssim 2\epsilon/(\pi \sqrt{N}) \).
Therefore, in order to preserve quadratic speedup in the presence of coherent errors,
we must suppress the perturbation strength to scale as \( \mathcal{O}(1/\sqrt{N}) \).

\section{Analytical verification of upper bound}

\subsection{ Time-Independent errors}

We now consider a time-independent coherent perturbation 
of the form \( K = \gamma Y \), where $\gamma$ represents 
 the error strength and $Y$ is defined by
\begin{eqnarray}
Y= I^{(1)}\otimes\cdots \otimes \sigma^{(i)}_y\otimes \cdots \otimes I^{(N)}.
\end{eqnarray}
$\sigma^{(i)}_y=-i|0\rangle \langle1|+i|1\rangle \langle0|$ is the Pauli-$y$ matrix acting on the $i$th qubit, where $|0\rangle=(1, 0)^\top$ and $|1\rangle=(0, 1)^\top$.
This operator generates local rotations of the atom around the \(y\)-axis on the Bloch sphere.

In this case, the norm \(\|K\|= \gamma \) is constant in time.
Substituting it into the obtained inequality yields
\begin{align}
\| |\psi(T)\rangle-  |\phi(T)\rangle\| &\leq \int_0^T \gamma \exp\left( \int_s^T \gamma\ du \right) ds \nonumber \\
&= \gamma \int_0^T e^{\gamma (T - s)} ds \nonumber \\
&= e^{\gamma T} - 1.
\end{align}
This upper bound depends only on the product \( \gamma T \), 
and it characterizes the worst-case accumulation of coherent error over time. 
The result shows that the deviation grows exponentially with time, 
which is consistent with coherent unitary drift caused by a systematic error term.
This expression provides a useful worst-case estimate 
when the perturbation is constant and aligned with a fixed generator such as \( Y \).
 It also serves as a reference point for evaluating the effect of time-dependent errors.

\subsection{Time-dependent errors}

We next consider a time-dependent coherent perturbation of the form \( K(t) = \gamma \sin(\omega t) Y \),
where \( \omega \) denotes the oscillation frequency.
Assuming that the total evolution time \( T \) is an integer multiple of the modulation period,
i.e., \( T = N \pi/\omega \) for some integer \( N \), 
the Grönwall-type inequality can be evaluated explicitly by exploiting the periodic structure of the perturbation.
Noting that \( \|K(s)\| = \gamma |\sin(\omega s)| \) and 
\begin{align}
\int_s^T \|K(u)\| du &= \gamma \int_s^T |\sin(\omega u)| du \nonumber \\
&\leq \gamma (T - s).
\end{align}
This upper bound allows us to factor the exponential term as
\begin{align}
\exp\left( \int_s^T \|K(u)\| du \right) 
\leq e^{\gamma (T - s)} = e^{\gamma T} e^{-\gamma s}.
\end{align}
Substituting this result into the Grönwall-type inequality yields
\begin{eqnarray}
\| |\psi(T)\rangle - |\phi(T)\rangle \| \leq \gamma e^{\gamma T} \int_0^T |\sin(\omega s)| e^{-\gamma s} ds.
\end{eqnarray}

The integral can be evaluated by partitioning the time interval into \( N \) complete periods:
\begin{eqnarray}
\int_0^T |\sin(\omega s)| e^{-\gamma s} ds = N \int_0^{\pi/\omega} |\sin(\omega s)| e^{-\gamma s} ds. \nonumber \\
\end{eqnarray}
Using the substitution \( \theta = \omega s \), the integral over one period becomes
\begin{align}
\int_0^{\pi/\omega} |\sin(\omega s)| e^{-\gamma s} ds 
&= \frac{1}{\omega} \int_0^{\pi} \sin(\theta) e^{-\frac{\gamma}{\omega} \theta} d\theta \nonumber \\
&= \frac{1}{\omega} \cdot \frac{1 + e^{-\frac{\gamma \pi}{\omega}}}{\left( \frac{\gamma^2}{\omega^2} + 1 \right)}.
\end{align}
Combining these expressions, we arrive at the compact result:
\begin{eqnarray}
\| |\psi(T)\rangle - |\phi(T)\rangle \| 
\leq C(\gamma, \omega) \cdot \gamma T \, e^{\gamma T},
\end{eqnarray}

where 
\begin{eqnarray}
C(\gamma, \omega)
:= \frac{\omega^2}{\pi \left(\gamma^2+  \omega^2 \right)} \left(1 + e^{-\frac{\gamma \pi}{\omega}}\right)
\end{eqnarray}

This expression represents the combined effects of the error amplitude, 
oscillation frequency, and evolution time on the deviation norm.
It provides a clear quantitative criterion for analyzing how time-dependent coherent errors accumulate in the system.
Interestingly, the analytic bound for the time-dependent coherent error 
is consistently larger than that for the time-independent case.
This arises because, while the instantaneous value of the time-dependent perturbation oscillates
and averages to a smaller effective strength,
the Grönwall-type inequality exponentially amplifies early-time contributions 
and linearly accumulates their overall effect over time.

Figure 1(a) and 2(b) illustrate the frequency dependence of the Grönwall-type bound for time-dependent coherent errors.
Both figures use the same error strength $\gamma = 0.15$, but different oscillation frequencies $\omega$.
In Fig.1 (a) ($\omega = \pi$), the bound for the time-dependent case remains relatively 
close to that of the time-independent case over the considered time range.
In contrast, in Fig.2 (b) ($\omega = \pi/4$), the time-dependent bound becomes  larger and exceeds the time-independent bound even at moderate evolution times.
These results implies that the frequency of coherent control errors is a 
critical factor in determining their impact on quantum algorithm performance.

\begin{figure}[tb]
\centering
\includegraphics[width=0.8\linewidth]{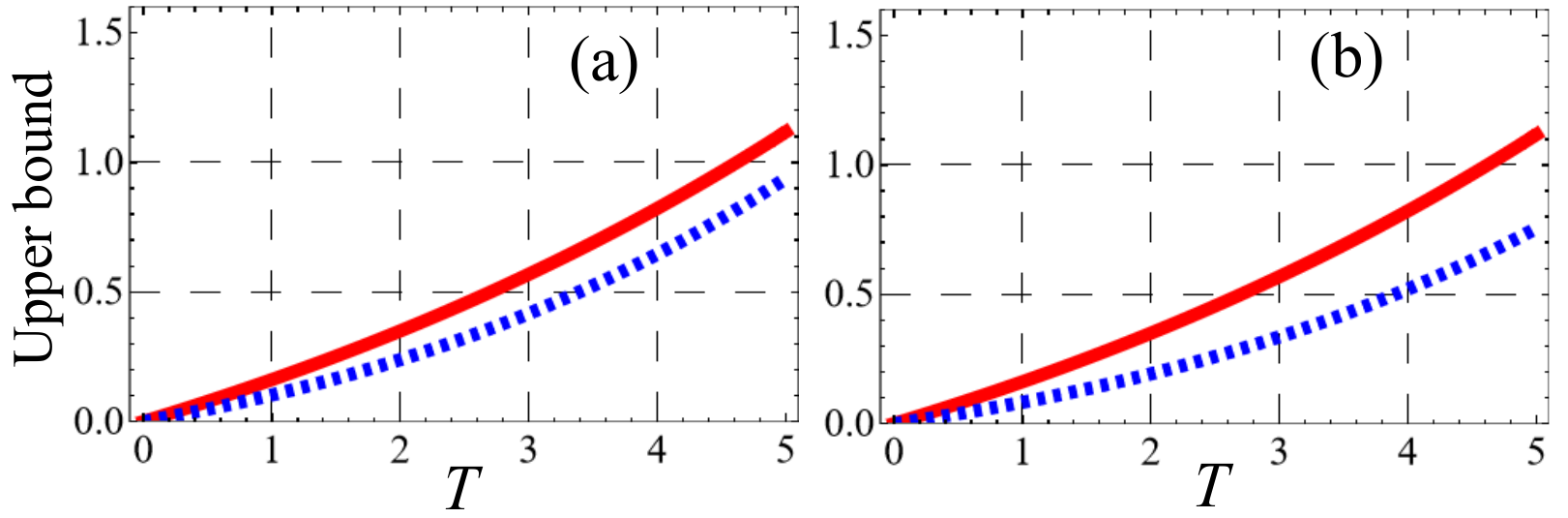}
\caption{
Comparison of the analytic upper bounds 
for time-dependent error (red solid line) and time-independent one (blue dotted line).
Both figures use the same error strength $\gamma = 0.15$.
Figure 1(a) shows the case of $\omega = \pi$, where the time-dependent bound remains relatively close to the time-independent bound.
Figure 1(b) shows the case of $\omega = \pi/4$, where the difference between the two plots becomes larger. }
\end{figure}

\section{Conclusion}
In this paper, we have investigated the influence of control errors on quantum state evolution 
by comparing the solutions of two Schrödinger equations; one governed by an ideal Hamiltonian \( H(t) \), 
and the other by a perturbed Hamiltonian \( H(t) + K(t) \). 
By deriving a Grönwall-type inequality for the norm of the state difference, 
we obtained a rigorous upper bound on the deviation induced by the perturbation, 
which depends on the operator norm of the error term \( K(t) \).
This bound enables the analysis of large-scale quantum systems without the need to simulate their full dynamics, and its exponential structure yields qualitative insight into the cumulative effects of perturbations over time.

We applied this framework to evaluate the performance of Grover's search algorithm under errors.
By relating the deviation bound to the algorithm's success probability, 
we derived a sufficient condition on the perturbation strength to ensure a given performance level.
Specifically, we found that to preserve a success probability of at least \( 1 - \epsilon \), 
the error strength must scale as \( \gamma = \mathcal{O}(1/\sqrt{N}) \).
This result elucidates a fundamental trade-off between speed and robustness in quantum search protocols.

Our analysis demonstrates that norm-based inequalities can serve as powerful tools 
for assessing and designing quantum algorithms under realistic control limitations.
Future work may extend this approach to incorporate dissipative dynamics, adaptive control strategies,
or feedback mechanisms, further bridging the gap between idealized quantum algorithms and practical quantum technologies.

\begin{acknowledgments}
This work was supported by MEXT Quantum Leap Flagship Program Grant JPMXS0120351339.
\end{acknowledgments}

\bibliographystyle{plain}

\begin{thebibliography}{9}


\bibitem{Nielsen}
M. A. Nielsen and I. L. Chuang,
{\it Quantum Computation and Quantum Information},
Cambridge University Press (2000).

\bibitem{Grover}
L. K. Grover,
A fast quantum mechanical algorithm for database search,
arXiv:quant-ph/9605043 (1996).

\bibitem{Farhi}
E. Farhi and S. Gutmann,
Analog analogue of a digital quantum computation,
Phys. Rev. A {\bf 57}, 2403 (1998).


\bibitem{Roland}
J. Roland and N. J. Cerf, 
Quantum search by local adiabatic evolution,
 Phys. Rev. A {\bf 65}, 042308 (2002).


 \bibitem{Biham}
O. Biham, D. Shapira, and Y. Shimoni,  
Analysis of Grover's quantum search algorithm as a dynamical system,
Phys. Rev. A {\bf 68}, 022326 (2003).


\bibitem{error1}
A. M. Childs,  E. Farhi, and J. Preskill,
Robustness of adiabatic quantum computation,
Phys. Rev. A {\bf 65}, 012322 (2001).

\bibitem{error2}
J. Roland and N. J. Cerf,
Noise resistance of adiabatic quantum computation using random matrix theory,
	Phys. Rev. A {\bf 71}, 032330 (2005).


\bibitem{error3}
K. C. Young, R. B.-Kohout, and D. A. Lidar,
Adiabatic quantum optimization with the wrong Hamiltonian,
Phys. Rev. A 88, 062314 (2013).


\bibitem{error4}
S. Mandrà, G. G. Guerreschi, and A. A.-Guzik,
Adiabatic quantum optimization in the presence of discrete noise: Reducing the problem dimensionality,
Phys. Rev. A {\bf 92}, 062320 (2015).

\bibitem{error5}
S. Muthukrishnan, T. Albash, D. A. Lidar,
Sensitivity of quantum speedup by quantum annealing to a noisy oracle,
Phys. Rev. A {\bf 99}, 032324 (2019).

\bibitem{error6}
A. Pearson, A. Mishra, I. Hen, and D. A. Lidar,
Analog Errors in Quantum Annealing: Doom and Hope,
	npj Quantum Information {\bf 5}, 107 (2019).

\bibitem{QEC1} 
P. W. Shor, 
Scheme for reducing decoherence in quantum computer memory, 
Phys. Rev. A {\bf 52}, R2493 (1995).

\bibitem{QEC2} 
A. M. Steane, 
Error correcting codes in quantum theory, 
Phys. Rev. Lett. {\bf 77}, 793 (1996).

\bibitem{DD1} 
L. Viola and S. Lloyd, 
Dynamical suppression of decoherence in two-state quantum systems, 
Phys. Rev. A {\bf 58}, 2733 (1998).

\bibitem{DD2} 
L. Viola, E. Knill, and S. Lloyd, 
Dynamical decoupling of open quantum systems, 
Phys. Rev. Lett. {\bf 82}, 2417 (1999).



\end{thebibliography}


\end{document}